\documentclass[onecolumn,aps,notitlepage,showpacs,superscriptaddress]{revtex4-2}
\usepackage{hyperref}
\usepackage{amsmath}
\usepackage{graphicx}
\usepackage{slashed}
\usepackage{xcolor}
\RequirePackage[pagewise,mathlines]{lineno}

\definecolor{PTturquoise}{RGB}{0,110,110}

\begin{document}
\title{Quantum Correlations Between Two Giant Dipole Resonances in Ultra-Peripheral Collisions}

\author{Albandry Alshammri}
    \affiliation{Center for Nuclear Research, Physics Department, Kent State University, Kent, OH 44242, USA}
    \affiliation{Physics Department, College of Science, Northern Border University, Arar 91431, Saudi Arabia}
\author{Ashik Ikbal Sheikh}
   % \email{ashikhep@gmail.com}
    \affiliation{Center for Nuclear Research, Physics Department, Kent State University, Kent, OH 44242, USA}
     \affiliation{ International Institute for Sustainability with Knotted Chiral Meta Matter (WPI-SKCM$^2$), Hiroshima University, Higashi-Hiroshima 739-8511, Japan}
\author{Declan Keane}
    \affiliation{Center for Nuclear Research, Physics Department, Kent State University, Kent, OH 44242, USA}
\author{Prithwish Tribedy}
    \affiliation{Physics Department, Brookhaven National Laboratory, Upton, NY 11973, USA}
    \author{Gang Wang}
    \affiliation{Department of Physics and Astronomy, University of California, Los Angeles, CA 90095, USA}
    \author{Zhangbu Xu}
    \affiliation{Center for Nuclear Research, Physics Department, Kent State University, Kent, OH 44242, USA}
    \affiliation{Physics Department, Brookhaven National Laboratory, Upton, NY 11973, USA}
    \author{Jie Zhao}
    \affiliation{Key Laboratory of Nuclear Physics and Ion-beam Application (MOE), Institute of Modern Physics, Fudan University, Shanghai, 200433, China}
\affiliation{Shanghai Research Center for Theoretical Nuclear Physics, NSFC and Fudan University, Shanghai, 200438, China}

\date{\today}

\begin{abstract}

Mutual Coulomb dissociation in ultra-peripheral heavy-ion collisions is dominated by Giant Dipole Resonance (GDR) excitation, whose subsequent decays produce forward neutrons. Standard descriptions treat the two GDRs excited in opposite nuclei as statistically independent, leading primarily to a quadrupolar azimuthal correlation determined by the common photon-polarization axis. In this work, we investigate the possibility that the two GDRs are instead prepared with a correlated quantum phase during the overlap of the electromagnetic pulses. We develop a phase-sensitive GDR decay model and implement it in a modified version of the \texttt{$n_{0}^{0}n$} Monte Carlo generator, allowing a quantitative comparison among three competing mechanisms: the weak dipolar modulation from electromagnetic recoil, the quadrupolar correlation from independent polarized GDR decays, and the phase-sensitive dipolar modulation associated with the proposed phase-lock scenario. For neutron pairs with small relative emission-time differences, the correlated-phase scenario produces a pronounced first-harmonic modulation in the two-neutron relative-azimuth distribution. Averaging over all relative emission times substantially reduces this modulation. Within the adopted energy-dependent angular model, the time-selected dipolar coefficient increases in magnitude with pair-mean neutron kinetic energy. These results identify neutron relative-azimuth correlations and their energy dependence as experimentally accessible observables for probing quantum phase correlations between the two GDR excitations. Such phase-sensitive correlations could also provide a basis for future tests of quantum entanglement between spatially separated GDR excitations at RHIC and the LHC.
\end{abstract}
\maketitle

\section{Introduction}

\begin{figure}[hpbt]
  \centering
  \includegraphics[width=1.02\linewidth]{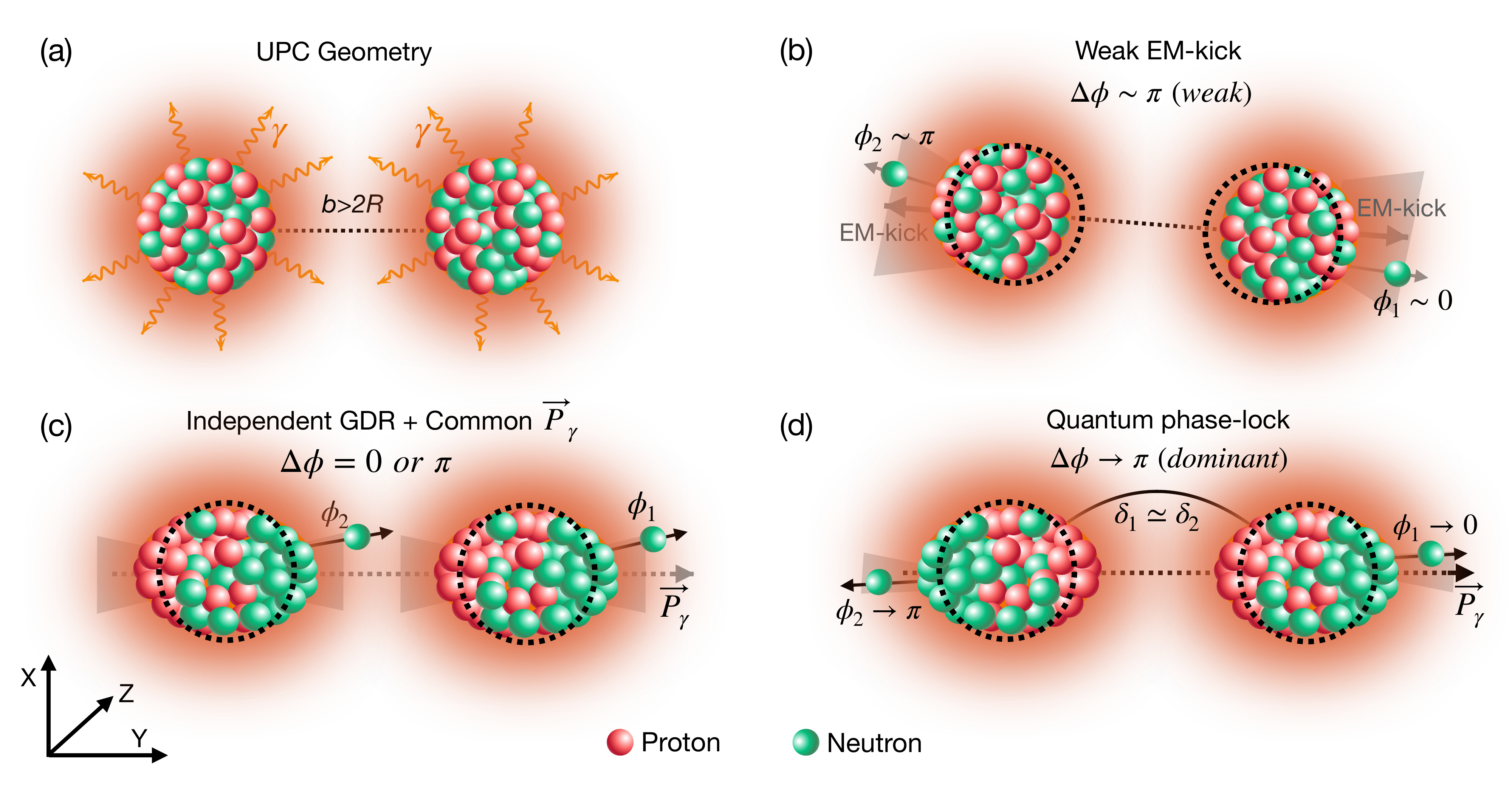}
\caption{Schematic illustration of the mechanisms contributing to two-neutron azimuthal correlations in mutual giant dipole resonance (GDR) excitation in ultra-peripheral heavy-ion collisions. The impact parameter $\vec{b}$ lies in the transverse $X$--$Y$ plane, while the $Z$ axis denotes the beam direction. 
(a) UPC geometry, in which the two nuclei pass at $b>2R$ and interact predominantly through their electromagnetic fields. 
(b) Electromagnetic recoil: the two nuclei receive small and oppositely directed transverse momentum kicks, producing a weak back-to-back dipolar correlation. 
(c) Independent polarized GDR decays: the common photon-polarization axis aligns the GDR oscillations in the two nuclei and produces a quadrupolar neutron correlation with comparable preference for $\Delta\phi\simeq0$ and $\pi$. 
(d) Phase-lock scenario: the two GDRs are prepared with a correlated phase, and phase-sensitive interference in their neutron decays generates an additional first-harmonic modulation that preferentially enhances back-to-back pairs. The shaded sectors indicate the schematic directions of enhanced neutron emission and are not intended to represent quantitative angular distributions.}
\label{fig1:gdr_cartoon}
\end{figure}

Photon-induced reactions in relativistic heavy-ion collisions provide a unique laboratory for studying quantum electrodynamics and nuclear structure under extreme electromagnetic fields. At the Relativistic Heavy Ion Collider (RHIC) and the Large Hadron Collider (LHC), ultra-peripheral collisions (UPCs), in which the impact parameter exceeds approximately the sum of the nuclear radii, strongly suppress hadronic interactions while allowing the two nuclei to interact through their intense Lorentz-contracted electromagnetic fields. As illustrated schematically in Fig.~\ref{fig1:gdr_cartoon}(a), these fields may be described in terms of quasi-real equivalent photons within the Equivalent Photon Approximation (EPA)~\cite{BertulaniBaur1988,Baltz2008,Brandenburg:2022tna}. UPCs therefore provide a clean environment in which photon-induced nuclear reactions can be investigated with minimal contamination from strong interactions.

One of the most important UPC channels is mutual Coulomb dissociation (MCD)~\cite{BaurBertulani1986,HarakehWoude2001}, in which each nucleus absorbs one or more photons emitted by its collision partner. For heavy nuclei such as $^{197}$Au and $^{208}$Pb, the excitation is dominated by the isovector Giant Dipole Resonance (GDR), a collective oscillation of protons against neutrons with a characteristic excitation energy of $E_\gamma\sim13$--15 MeV. The subsequent GDR decay produces forward neutrons that are detected experimentally using Zero Degree Calorimeters (ZDCs)~\cite{ZDC2003}. Because the neutron multiplicity and energy can be measured with high efficiency, mutual GDR excitation has become one of the standard probes of UPC dynamics. Existing descriptions, including STARlight-based implementations~\cite{STARlight2017}, treat the two GDR excitations and their decays as statistically independent apart from their common collision geometry.

Recent UPC measurements~\cite{STAR:2019wlg,STAR:2022wfe,ALICE_upc_spin_qm,BrandenburgKleinXu2025Review,BrandenburgDuanTu2025,Klein2026} have demonstrated that the common electromagnetic field of a single ion-ion encounter can leave measurable polarization and coherence imprints on the final state. In particular, the linear polarization ($\vec{P}_{\gamma}$) of equivalent photons is aligned with the impact-parameter direction $\vec{b}$, defining a preferred transverse axis on an event-by-event basis. In vector-meson photoproduction~\cite{STAR:2022wfe,ALICE_upc_spin_qm,STAR:2025wpi}, this polarization gives rise to characteristic azimuthal anisotropies, illustrating that UPC observables can retain information about the geometry of the initial electromagnetic field.

These considerations naturally lead to the question of how the common photon field influences mutual GDR excitation. As summarized schematically in Fig.~\ref{fig1:gdr_cartoon}, the neutron relative-azimuth distribution receives contributions from three distinct mechanisms. The first is a weak dipolar correlation arising from the electromagnetic recoil of the two nuclei [Fig.~\ref{fig1:gdr_cartoon}(b)]. The second is a quadrupolar correlation generated by statistically independent but polarized GDR decays aligned with the common photon-polarization axis [Fig.~\ref{fig1:gdr_cartoon}(c)]. Neither mechanism, however, produces a pronounced first-harmonic modulation in the neutron relative azimuth. This motivates the central question of the present work: can the two GDRs instead be prepared with a correlated quantum phase during the brief overlap of the electromagnetic pulses? If so, the subsequent neutron decays may retain signatures of this phase correlation. As illustrated in Fig.~\ref{fig1:gdr_cartoon}(d), such a phase-lock scenario would generate a strong first-harmonic modulation in the neutron relative-azimuth distribution that is qualitatively distinct from the conventional recoil and independent-decay mechanisms.

In this work, we investigate this possibility by implementing a phase-sensitive GDR decay model in a modified version of the \texttt{nOOn} Monte Carlo generator~\cite{Broz:2019kpl}. The framework allows a quantitative comparison among the three competing mechanisms under realistic UPC conditions. For pairs selected to have small relative emission-time differences, the phase-lock scenario produces a pronounced first-harmonic modulation, while the reference mechanisms exhibit a weak dipole or a quadrupole. The predicted strength also depends on the prescribed energy-dependent neutron angular anisotropy. These observables provide a concrete experimental strategy for testing correlated quantum phases in mutual GDR excitation at RHIC and the LHC, and more broadly illustrate how photon-induced reactions in relativistic heavy-ion collisions can probe coherence phenomena between spatially separated nuclear systems.

\section{Theoretical Framework}
\label{theory}

Direct photonuclear measurements show that the retention of polarization information is not limited to vector mesons~\cite{HarakehWoude2001,Heyde:2010ng}. Experiments using linearly polarized $\gamma$ rays in the GDR energy range have established that neutron emission from GDR decay exhibits a pronounced $\cos2\phi$ dependence relative to the photon polarization axis. Horikawa \emph{et al.}~\cite{Horikawa:2014PLB} observed clear quadrupole anisotropies in neutron emission from $^{197}$Au, $^{127}$I, and natural Cu. Kirihara \emph{et al.}~\cite{Kirihara:2019JNST} demonstrated that the prompt, direct neutron component from polarized 16.6~MeV photons carries a strong angular modulation, while the evaporative tail remains nearly isotropic. Similar polarization-driven patterns were mapped across the GDR region in $^9$Be($\gamma$,n) reactions~\cite{Kikuchi:2018PRC} and confirmed at NewSUBARU~\cite{Miyamoto:2018PFR}. These results show that the GDR neutron-emission pattern can faithfully encode the incident photon polarization, providing a classical baseline against which quantum-coherent effects may be compared.

To probe this possibility, we examine the azimuthal correlation of forward-backward neutrons emitted from the two nuclei. This observable receives contributions from multiple mechanisms. Classical Coulomb repulsion induces a weak dipolar modulation, $1 + a_1 \cos\Delta\phi$ with $a_1 < 0$~\cite{Zhao:2022dac}, due to opposite transverse momentum kicks imparted to the nuclei by the electromagnetic fields. Independent GDR decays in the exchanged photon polarization field generate a robust quadrupolar structure, $1 + a_2 \cos2\Delta\phi$~\cite{HarakehWoude2001,KleinNystrand2000}. Neither mechanism, however, produces a pronounced first-harmonic term. In contrast, if the two GDRs share a correlated initial quantum phase, the interference between direct and resonant decay pathways can generate a pronounced first-harmonic dipolar modulation 
that survives averaging over photon polarization. Critically, this first-harmonic component is sensitive to the phase correlation between the two GDRs. Its strength depends on how well this correlation is preserved against effects such as differences in the neutron emission times and dephasing due to mismatched GDR frequencies. Such a correlated phase could arise from an entangled two-nucleus state, although observation of the first-harmonic signal alone would not establish entanglement. One possible microscopic mechanism for generating such correlations, involving correlated E1/M1 excitation amplitudes in the two nuclei, is discussed below. The azimuthal correlations in mutual GDR excitation can therefore be understood as arising from three distinct contributions:
\begin{enumerate}

\item a weak dipole from electromagnetic recoil;  
\item a quadrupole from polarization-driven but independent GDR decays; and  
\item a phase-sensitive dipole that appears only if the two GDRs share a correlated quantum phase. 
\end{enumerate}

We analyze these three contributions in the following subsections.

\subsection{The EM-Kick and Recoil-Induced Dipole Background}
\label{emkick}

A classical source of azimuthal anisotropy arises from the electromagnetic recoil (EM kick) imparted to the two nuclei during an ultra-peripheral collision. Because the Lorentz-boosted electromagnetic fields of each nucleus act on its collision partner, the two ions experience strong, oppositely directed transverse forces in the instant of closest approach. As illustrated in Fig.~\ref{fig1:gdr_cartoon}(b), this classical impulse pushes the nuclei apart in the transverse plane. This mechanism was recently quantified in Monte Carlo simulations by Zhao \textit{et al.}~\cite{Zhao:2022dac}, who showed that even though the fields act over a very short time, they impart a measurable transverse momentum shift.

In relativistic UPCs, the colliding nuclei are extremely Lorentz contracted, and the interaction time is tiny (\(\Delta t \sim 2R/(\gamma v_z)\))~\cite{Zhao:2022dac}. Over this interval, the electric and magnetic fields reach magnitudes of  
\(eE \sim eB \sim 0.2\,m_\pi^2\) for Au+Au at \(\sqrt{s_{_{NN}}} = 200\) GeV (with \(b \sim 20\) fm)~\cite{Zhao:2022dac}.  
Using the Li\'{e}nard--Wiechert formulation (Eqs. (2)--(3) in Ref.~\cite{Zhao:2022dac}), the total impulse is
\[
\Delta \mathbf{p}_T = \int dt\, q(\mathbf{E} + \mathbf{v}\times \mathbf{B}).
\]
The analytic estimate in Ref.~\cite{Zhao:2022dac} gives a transverse momentum kick of approximately 2~MeV/$c$ per nucleon. Although the Lorentz force acts directly only on the protons, the interaction time is much shorter than the characteristic timescale for nuclear breakup. Consequently, the nucleus responds coherently as a strongly bound object, and the recoil momentum is shared among all nucleons through the strong nuclear interaction. 

Although this recoil momentum is small compared with the typical GDR neutron momentum (\(\sim 100\) MeV)~\cite{Zhao:2022dac}, it was demonstrated that the EM kick generates a clear back-to-back correlation between neutrons emitted from opposite nuclei (Fig. 2 in Ref.~\cite{Zhao:2022dac}) in the absence of any other physics effect that would generate larger azimuthal anisotropy. The above simulations show that the resulting first-harmonic modulation in the neutron--neutron relative azimuth can be described by
$1 + a_1 \cos\Delta\phi$ 
with \(a_1 \simeq -0.044\pm 0.013\) for Au+Au UPCs at \(\sqrt{s_{NN}}=200\) GeV (Fig. \ref{fig2:delphi}(a), red curve).  
The negative sign arises from the nuclei being pushed apart: neutrons are preferentially emitted in opposite azimuthal directions. This study also found that the EM kick depends on the impact parameter: the recoil is smallest in the most peripheral $1n1n$ events, in which one neutron is emitted from each nucleus, and increases gradually in $XnXn$ events, in which multiple neutrons are emitted from each nucleus; the latter preferentially sample smaller impact parameters and somewhat higher photon energies~\cite{Zhao:2022dac}. Even in the latter case, the resulting dipole amplitude remains modest, always \( |a_1| \lesssim 0.05 \) because the momentum kick is always more than an order of magnitude smaller than the neutron total momentum.

\subsection{Independent Polarized GDR Decays}
\label{single_gdr}
The second contribution arises from the intrinsic angular distribution of neutrons emitted in the decay of a linearly polarized GDR. In a UPC, the equivalent-photon electric field is linearly polarized approximately along the impact-parameter direction $\vec{b}$~\cite{STAR:2019wlg,STAR:2022wfe}. Consequently, each nucleus is preferentially excited into a GDR with its oscillation axis aligned with $\vec{b}$, as illustrated in Fig.~\ref{fig1:gdr_cartoon}(c). This GDR corresponds to a collective oscillation of the proton and neutron fluids within the nucleus, and its subsequent neutron decay retains information about the oscillation axis, giving rise to an anisotropic angular distribution.

The excited GDR evolves as a damped collective doorway state,
\begin{equation}
c_0(t)=c_0(0)e^{-\Gamma t/2}e^{-i\omega t},
\label{eq:gdr_time}
\end{equation}
where $\omega$ is the GDR oscillation frequency and $\Gamma$ is its decay width. Equation~(\ref{eq:gdr_time}) describes the time evolution of the polarized GDR prior to neutron emission. As discussed in the following subsection, the finite lifetime $\Gamma^{-1}$ sets the characteristic timescale over which the GDR can retain phase information.

Strong experimental evidence exists that neutron emission following GDR excitation by linearly polarized photons exhibits a pronounced $\cos2\phi$ anisotropy. Using nearly 100\% linearly polarized laser Compton--scattered photons in the GDR energy range (12--16 MeV), Horikawa \textit{et al.}~\cite{Horikawa:2014PLB} measured neutron angular distributions for $^{197}$Au, $^{127}$I, and Cu, finding
\[
\frac{dN}{d\phi}\propto a+b\cos2\phi,
\]
with amplitudes consistent with electric-dipole (E1) excitation as predicted by Agodi's formalism~\cite{agodi1957polarization}. Subsequent high-precision measurements by Kirihara \textit{et al.}~\cite{Kirihara:2019JNST} using 16.6 MeV monoenergetic polarized photons on $^{197}$Au demonstrated that the direct neutron-emission component (above $\sim4$ MeV) carries a strong quadrupole pattern, whereas the evaporative component remains nearly isotropic. Similar polarization-dependent angular distributions have also been observed in the $^{9}$Be$(\gamma,n)$ reaction~\cite{Kikuchi:2018PRC} and confirmed at NewSUBARU by Miyamoto \textit{et al.}~\cite{Miyamoto:2018PFR}. These measurements establish that neutron emission from a polarized GDR preserves the incident photon polarization and provides a well-defined classical baseline.

Motivated by these experimental results, we describe the neutron angular distribution from each independently excited nucleus by
\begin{equation}
\frac{dN}{d\phi}\propto1+a_2\cos2\phi,
\label{eq:single_gdr}
\end{equation}
where $\phi$ is measured relative to the photon polarization axis (equivalently, the impact-parameter direction), and $a_2$ denotes the quadrupole anisotropy parameter determined from polarized photo-neutron measurements.
The physical meaning of Eq.~(\ref{eq:single_gdr}) is illustrated schematically in Fig.~\ref{fig1:gdr_cartoon}(c). Because the GDR oscillates along the photon-polarization axis, neutron emission preferentially occurs parallel and antiparallel to that axis. These two directions are physically equivalent since the dipole oscillation is invariant under a $180^\circ$ reversal. Consequently, the neutron angular distribution is $\pi$-periodic and is naturally described by the second harmonic, $\cos2\phi$, rather than the first harmonic.

When the two nuclei in a UPC decay independently but share the same event-by-event polarization axis defined by $\vec{b}$, the convolution of the two single-nucleus distributions naturally yields a quadrupolar correlation in the relative azimuth,
\begin{equation}
C_{\rm pol}(\Delta\phi)=1+\frac{1}{2}a_2^2\cos2\Delta\phi.
\label{eq:gdr2}
\end{equation}
Equation~(\ref{eq:gdr2}) represents the expected two-neutron correlation in the independent-decay picture. Although the two nuclei share the same polarization axis established by the common electromagnetic field, their decays are otherwise statistically independent. As a result, the relative-azimuth distribution is dominated by the quadrupole term, while no first-harmonic ($\cos\Delta\phi$) modulation is generated. This quadrupolar correlation therefore serves as the classical baseline against which the phase-sensitive mechanism introduced in the following subsections may be compared~\cite{Baur:2003ar,crawford6star}.

\subsection{Single-Nucleus Interference: Direct and Doorway Pathways}

To go beyond the independent-decay picture, we now consider the neutron-emission amplitude from a single excited nucleus. As discussed in Sec.~\ref{single_gdr}, the polarized GDR evolves as a damped doorway state according to Eq.~(\ref{eq:gdr_time}). The corresponding resonant neutron-emission amplitude may therefore be written as
\begin{equation}
M_{\rm res}(t,\phi)
=
B\,e^{-\Gamma t/2}e^{-i\omega t}\cos\phi,
\label{eq:Mres}
\end{equation}
where $B$ is the resonant amplitude and the angular factor $\cos\phi$ reflects the polarization of the GDR established by the incident linearly polarized photon.

In addition to this resonant doorway contribution, there is in general a nonresonant direct contribution in which the photon excites continuum states that emit neutrons on timescales much shorter than $\Gamma^{-1}$. Near the GDR peak this direct contribution varies slowly with energy and may be approximated by
\begin{equation}
M_{\rm direct}(t)\approx A.
\end{equation}

Because both pathways populate the same final neutron channel, their amplitudes must be added coherently rather than at the cross-section level. The total single-nucleus decay amplitude is therefore
\begin{equation}
M_i(t,\phi_i)
=
A_i
+
B_i e^{-\Gamma_i t_i/2}
e^{-i\omega_i t_i}
e^{i\delta_i}
\cos\phi_i,
\end{equation}
where $\delta_i$ denotes the relative phase between the direct and resonant doorway amplitudes.
This phase has a natural interpretation in the standard theory of resonance-background interference. The direct pathway carries a potential-scattering phase, whereas the doorway contribution carries the resonance phase associated with the Breit--Wigner propagator~\cite{Fano1961,Feshbach1958}. The effective phase difference may therefore be written schematically as
\begin{equation}
\delta(E)
=
\delta_{\rm resonance}(E)
-
\delta_{\rm potential}(E).
\end{equation}

Squaring the total amplitude gives
\begin{equation}
\label{phase-lock}
|M_i|^2
=
A_i^2
+
2A_iB_i
e^{-\Gamma_i t_i/2}
\cos(\omega_i t_i-\delta_i)
\cos\phi_i
+
B_i^2
e^{-\Gamma_i t_i}
\cos^2\phi_i .
\end{equation}
Equation~(\ref{phase-lock}) is an effective interference ansatz; the required microscopic transition amplitudes and their angular-momentum structure are not calculated here. It is the simplest form consistent with Eq.~(\ref{eq:single_gdr}): averaging over the emission time ($\Gamma e^{-\Gamma t}$) and over a phase $\delta_i$ that varies randomly from event to event removes the interference term and gives $1+a_2\cos2\phi$ with $a_2=B_i^2/(4A_i^2+B_i^2)$, while the isotropic direct term $A_i$ is what allows a first harmonic once the phases of the two nuclei are correlated. Equation~(\ref{phase-lock}) clearly separates the origins of the two azimuthal harmonics. The interference between the direct and doorway amplitudes generates the first-harmonic term proportional to $\cos\phi_i$, while the resonant doorway contribution produces the second harmonic through $\cos^2\phi_i$. Thus, already at the single-nucleus level, the dipolar component is intrinsically phase sensitive, whereas the quadrupolar component is determined solely by the polarized GDR geometry discussed in Sec.~\ref{single_gdr}.

\subsection{Correlated Phases in Mutual GDR Excitation}
\label{decoh}

The previous subsection considered the independent-decay limit. We now explore the possibility that the two GDRs are instead prepared with a correlated quantum phase during the overlap of the two electromagnetic pulses. In this case, the phase-sensitive interference terms derived in Eq.~(\ref{phase-lock}) need not average independently to zero. Their correlated contribution can survive averaging over the random event-by-event orientation of the photon-polarization axis and generate an additional first-harmonic modulation in the observable two-neutron relative-azimuth distribution, illustrated schematically in Fig.~\ref{fig1:gdr_cartoon}(d),

\begin{equation}
\frac{dN}{d\Delta\phi}
=
C_0
+
C_1\cos\Delta\phi
+
C_2\cos2\Delta\phi,
\end{equation}
where $\Delta\phi=\phi_1-\phi_2$. The coefficient $C_1$ inherits the phase-sensitive interference factor

\begin{equation}
C_1
\propto
\cos\!\left[\omega(t_1-t_2)-(\delta_1-\delta_2)\right],
\end{equation}
whereas $C_2$ is governed primarily by the doorway-strength terms and is much less sensitive to the relative phase.

Equation (10) immediately shows that a first-harmonic (dipolar) correlation can survive only if two conditions are satisfied. First, the two nuclei must be created with correlated phases so that $\delta_1-\delta_2$ remains small. Second, the neutron emissions must be prompt on the GDR oscillation timescale ($T_{\rm osc}$), {\it i.e.}, $|t_1-t_2|\ll T_{\rm osc}$, so that the oscillatory interference term does not average away. Under these conditions, the phase-sensitive contribution to the two-neutron distribution may be written phenomenologically as

\begin{equation}
\left(\frac{dN}{d\Delta\phi}\right)_{\rm phase}
=
1+c_1\cos\Delta\phi,
\end{equation}
which can be substantially larger than the weak recoil-induced dipole. For the phase convention considered here, this contribution preferentially enhances back-to-back neutron pairs, corresponding to $\Delta\phi\rightarrow\pi$, as illustrated in Fig.~\ref{fig1:gdr_cartoon}(d). The enhancement is probabilistic rather than exclusive: the phase-sensitive mechanism modifies the relative-azimuth distribution but does not require every event to exhibit an exactly back-to-back topology.

Two mechanisms suppress this first-harmonic contribution. The first is \emph{phase decoherence}: if the two nuclei are not prepared with the same phase ($\delta_1\neq\delta_2$), the dipolar term is reduced even for prompt neutron pairs. A convenient phenomenological parametrization introduces an event-by-event phase mismatch width $\sigma_\delta$, giving the suppression factor

\begin{equation}
\kappa_\delta
=
\exp(-\sigma_\delta^2/2).
\end{equation}

The second is \emph{time dephasing}: even if the initial phases are correlated, the first harmonic is suppressed when $|t_1-t_2|$ becomes comparable to or exceeds the GDR oscillation period. Combining these effects, the surviving first-harmonic strength for prompt neutron pairs may be summarized schematically as

\begin{equation}
C_1^{\rm prompt}
\propto
\kappa_\delta
e^{-\Gamma(t_1+t_2)/2}
\cos\omega(t_1-t_2),
\end{equation}
whereas for delayed neutron emission the first harmonic is strongly reduced or disappears altogether. By contrast, the quadrupolar coefficient $C_2$ survives independent phases and delayed emission over a much broader range of conditions. 

In the present work, the correlated phase is introduced phenomenologically. We do not attempt to identify its microscopic origin uniquely. One possible interpretation is that the two GDR excitations are jointly prepared in a two-body state carrying a correlated phase during the passage of the nuclei through the common electromagnetic field. Our objective is instead to determine the observable consequences of such correlated phases. If realized experimentally, the resulting first-harmonic neutron correlation would provide a direct signature that is absent in the conventional independent-decay picture. It is important to emphasize that the phase-lock description adopted here is phenomenological, while GDR excitation may involve multiple excitation channels, including E1 and M1 contributions~\cite{Horikawa:2014PLB,agodi1957polarization,Takeshita:2026xvn}. Recent investigations of coherence lengths and quantum correlations in nuclear radiative processes demonstrate that nuclear systems can, under suitable conditions, preserve phase information and even exhibit quantum entanglement in photon-- nucleus degrees of freedom~\cite{Mirzaee2025Coherence,SadeghiMirzaee2025EPJPlus}. Although these studies do not address mutual Coulomb excitation directly, they illustrate the broader plausibility of quantum-coherent behavior in nuclear dynamics. In mutual excitation, coherent correlations among E1 and M1 excitation amplitudes--well established in nuclear dipole–response studies and shown to exhibit overlapping strength and interference effects~\cite{Tamii2024E1M1, Sasaki2022FAM, Voinov2015Nd}—provide a plausible microscopic mechanism for generating the correlated two-nucleus states and offer a potential microscopic foundation for quantum entanglement in UPC-induced GDR excitations.

\section{Implementation in the $n^0_0n$ Event Generator}
The numerical simulation of mutual GDR de-excitation and the resulting neutron azimuthal correlations is performed using a heavily modified version of the (\texttt{$n_{0}^{0}n$}) Monte Carlo generator~\cite{Broz:2019kpl}. The default version of $n_{0}^{0}n$ provides a realistic description of UPC geometry and neutron multiplicities at RHIC and LHC energies, but it does not incorporate the effects of photon polarization or the phase-sensitive GDR dynamics developed in Sec.~\ref{theory}. In this section, we first briefly review the standard $n_{0}^{0}n$ framework and then describe the extensions implemented to include polarized GDR decay and quantum phase correlations.

\subsection{The Default $n_{0}^{0}n$ Framework}
The $n_{0}^{0}n$ program~\cite{Broz:2019kpl} is a C++/ROOT-based Monte Carlo generator that calculates the probability of electromagnetic dissociation (EMD) by integrating the equivalent-photon flux with the photonuclear cross sections. To ensure that only UPCs are selected, the generator suppresses hadronic interactions through a factor $\exp[-P_H(b)]$, where $P_H(b)$ is the mean number of nucleon--nucleon interactions at impact parameter $b$. This quantity is evaluated using the inelastic nucleon--nucleon cross section, the nuclear thickness function, and a Woods--Saxon nuclear density profile,
\begin{equation}
\rho_A(s)=\frac{\rho_0}{1+\exp\!\left[(s-R_{\rm WS})/d\right]},
\end{equation}
where $R_{\rm WS}$ and $d$ are the Woods--Saxon radius and diffuseness parameters, respectively.

The flux of quasi-real photons with energy $k$ at an impact parameter $b$ is calculated within the semi-classical Equivalent Photon Approximation (EPA)~\cite{Vidovic:1992ik,Broz:2019kpl,Zha:2018tlq},

\begin{equation}
\frac{d^3n(b,k)}{dk\,d^2\vec{b}}
=
\frac{Z^2\alpha\,k}{\pi^2\gamma^2}
\left[
K_1^2\!\left(\frac{kb}{\gamma}\right)
+
\frac{1}{\gamma^2}
K_0^2\!\left(\frac{kb}{\gamma}\right)
\right],
\label{flux}
\end{equation}
where $\alpha$ is the fine-structure constant, $\gamma$ is the Lorentz factor of the beam, and $K_0$ and $K_1$ are modified Bessel functions (natural units, $\hbar=c=1$).

For each event, the neutron multiplicity is sampled from a Poisson distribution of electromagnetic excitations, while the neutron kinetic energy is generated from the Evaluated Nuclear Data File (ENDF) tables~\cite{Kahler:2011hou}, which provide the energy spectra of photo-neutron reactions. In the default implementation, neutrons are emitted isotropically in the rest frame of the excited nucleus.

\subsection{Implementation of GDR Physics into $n_{0}^{0}n$}

To investigate mutual GDR excitation and the resulting neutron azimuthal correlations, we extend the isotropic neutron-emission model in the default $n_{0}^{0}n$ framework to incorporate the polarized GDR decay model developed in Sec.~\ref{theory}. In a UPC, the equivalent photons are linearly polarized approximately along the impact-parameter direction $\vec{b}$. Accordingly, the neutron angular distribution in the rest frame of each excited nucleus is generated according to
\begin{equation}
\frac{dN}{d\phi}\propto1+A_2\cos2\phi,
\end{equation}
where $\phi$ is measured relative to the photon-polarization axis. The quadrupole parameter $A_2$ is implemented as an explicit function of the neutron kinetic energy $E_n$, constrained by recent high-precision photo-neutron measurements using linearly polarized photon beams~\cite{THUONG2025139900}.

UPC events are generated by sampling the photon energy $k$ and impact parameter $b$ according to the EPA photon flux given by Eq.~\ref{flux}, while the orientation of $\vec{b}$ is randomized event by event. To reproduce the experimental trigger conditions, the photon energy is restricted by requiring coherent $\rho^0$ photoproduction at midrapidity ($|y|<1$),
$k=\frac{1}{2}M_{\rho^0}e^{\pm y}$. Coherent $\rho^0$ production defines the UPC event sample, while the GDR excitations arise from additional photon exchanges in the same event.

For the independent-decay scenario, each nucleus is generated according to the single-nucleus angular distribution above, and the two neutron emissions are sampled independently. Figure~\ref{fig2:delphi}(b) shows the simulated baseline for this independent-decay scenario in mutual Coulomb dissociation of Au+Au UPCs at $\sqrt{s_{NN}}=200$ GeV. In this example, neutrons are generated with a fixed kinetic energy of $E_n=7$ MeV and $A_2=0.5$, consistent with polarized photo-neutron measurements~\cite{THUONG2025139900}. A Fourier decomposition of the simulated two-neutron relative-azimuth distribution,
$f(\Delta\phi)
=
1+a_1\cos\Delta\phi
+a_2\cos2\Delta\phi$,
yields a pronounced quadrupole coefficient, $a_2=0.103\pm0.005$, while the first harmonic remains consistent with zero ($a_1=-0.008\pm0.005$). This confirms that independent polarized GDR decays naturally produce the expected $\cos2\Delta\phi$ correlation but essentially no dipolar component. Since this quadrupolar modulation is substantially stronger than the recoil-induced EM kick discussed in Sec.~\ref{emkick}, it provides the dominant classical contribution in the absence of quantum phase correlations.

The coherent phase-lock model described in Sec.~\ref{decoh} is incorporated into the modified $n_{0}^{0}n$ generator by replacing the statistically independent angular sampling of the two nuclei with the phase-sensitive single-nucleus probability of Eq.~\ref{phase-lock}, applied separately to each nucleus, while correlating their initial GDR phases, $\delta_1 \sim \delta_2$. This implementation introduces the temporal and phase correlations associated with a common initial GDR phase while leaving the UPC geometry, photon flux, neutron multiplicity, and neutron energy spectrum unchanged. Consequently, the resulting neutron pairs can carry the phase information established during the overlap of the two electromagnetic pulses, allowing the phase-sensitive first-harmonic modulation to be studied directly. The corresponding simulated two-neutron relative-azimuth distribution is shown in Fig.~\ref{fig2:delphi}(c), where the emergence of a pronounced first-harmonic modulation illustrates the characteristic signature of the phase-lock scenario. For Figs.~\ref{fig2:delphi}(c)--\ref{fig4:a1_en}, neutron pairs are further selected by $|t_1-t_2|<T_{\rm osc}/10$, where $T_{\rm osc}=2\pi/\omega$. This generator-level selection retains approximately 19.7\% of pairs and constrains the relative emission time rather than requiring both neutrons to be emitted early. The plotted result is therefore a conditional benchmark, not a time-inclusive experimental prediction.

\section{Results and Discussions}

\begin{figure}[!t]
  \centering
  \includegraphics[width=1.03\linewidth]{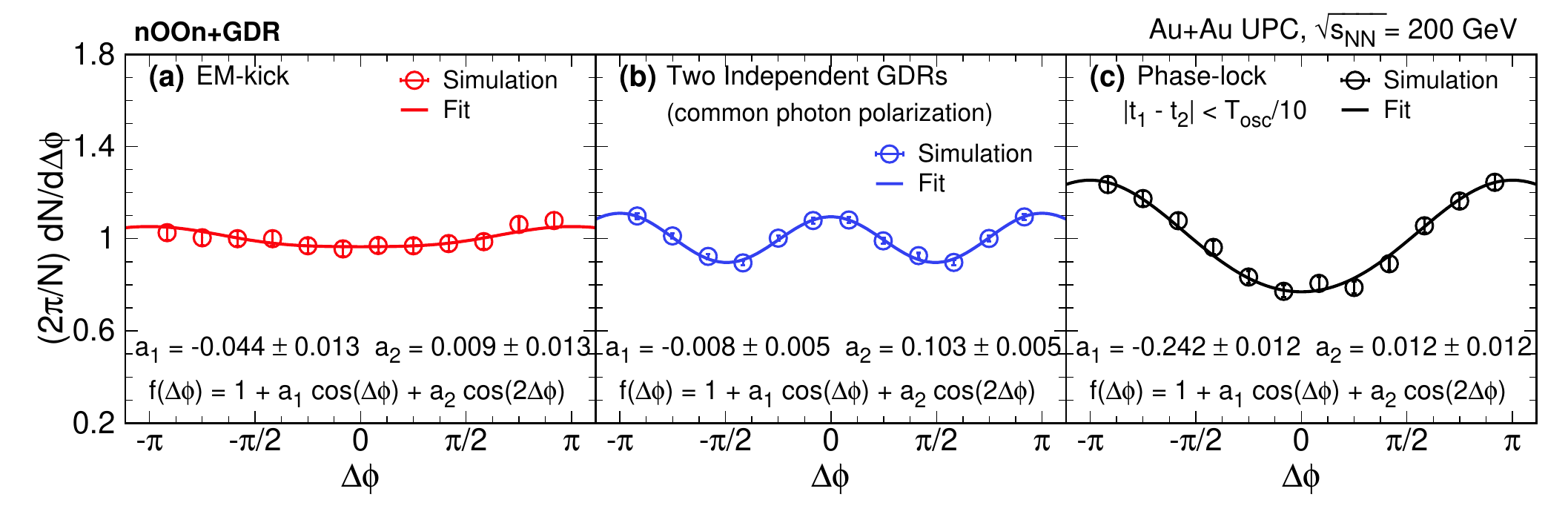}
\caption{Simulated two-neutron relative-azimuth distributions, $dN/d\Delta\phi$, for illustrative benchmark samples in Au+Au ultra-peripheral collisions at $\sqrt{s_{NN}}=200$ GeV, obtained with the modified \texttt{$n_{0}^{0}n$} generator. The distributions are fitted with the Fourier expansion $f(\Delta\phi)=1+a_1\cos\Delta\phi+a_2\cos2\Delta\phi$, where $\Delta\phi=\phi_1-\phi_2$. (a) Electromagnetic (EM) recoil generates only a weak dipolar modulation ($a_1=-0.044\pm0.013$). (b) Independent polarized GDR decays produce a pronounced quadrupolar correlation ($a_2=0.103\pm0.005$), while the dipolar component remains consistent with zero ($a_1=-0.008\pm0.005$). (c) The correlated-phase benchmark includes several surviving neutron-multiplicity classes and selects $|t_1-t_2|<T_{\rm osc}/10$, giving $a_1=-0.242\pm0.012$. The three panels correspond to different simulation scenarios and event selections.}
\label{fig2:delphi}
\end{figure}

Figure~\ref{fig2:delphi} summarizes the three mechanisms discussed in Sec.~\ref{theory} and implemented in the modified $n_{0}^{0}n$ generator. The corresponding two-neutron relative-azimuth distributions provide a direct comparison of the weak recoil-induced dipole, the quadrupolar correlation from independent polarized GDR decays, and the phase-sensitive dipole associated with the phase-lock scenario. Their Fourier coefficients quantify the relative strengths of the dipolar ($a_1$) and quadrupolar ($a_2$) components, allowing the different physical mechanisms to be clearly distinguished.

The three panels of Fig.~\ref{fig2:delphi} exhibit qualitatively different azimuthal structures. The EM-kick shown in Fig.~\ref{fig2:delphi}(a) produces only a weak first-harmonic modulation, with $a_1=-0.044\pm0.013$, reflecting the small transverse recoil imparted to the nuclei. Independent polarized GDR decays in Fig.~\ref{fig2:delphi}(b) instead generate a pronounced quadrupolar correlation ($a_2=0.103\pm0.005$), while the dipolar component remains consistent with zero ($a_1=-0.008\pm0.005$). This demonstrates that the shared photon-polarization axis alone cannot produce a first-harmonic correlation. In contrast, the phase-lock scenario in Fig.~\ref{fig2:delphi}(c) gives rise to a qualitatively different topology, characterized by a deep minimum at $\Delta\phi=0$ and maxima near $\Delta\phi=\pm\pi$, corresponding to a large dipolar coefficient of $a_1=-0.242\pm0.012$. This large coefficient characterizes the time-selected correlated-phase benchmark. Without the relative-emission-time selection, the same simulation yields $a_1=-0.0365\pm0.0054$. Because the three panels correspond to different simulation scenarios and event selections, their quantitative difference should not be interpreted as an isolated effect of phase locking.

\begin{figure}[!t]
  \centering
  \includegraphics[width=0.99\linewidth]{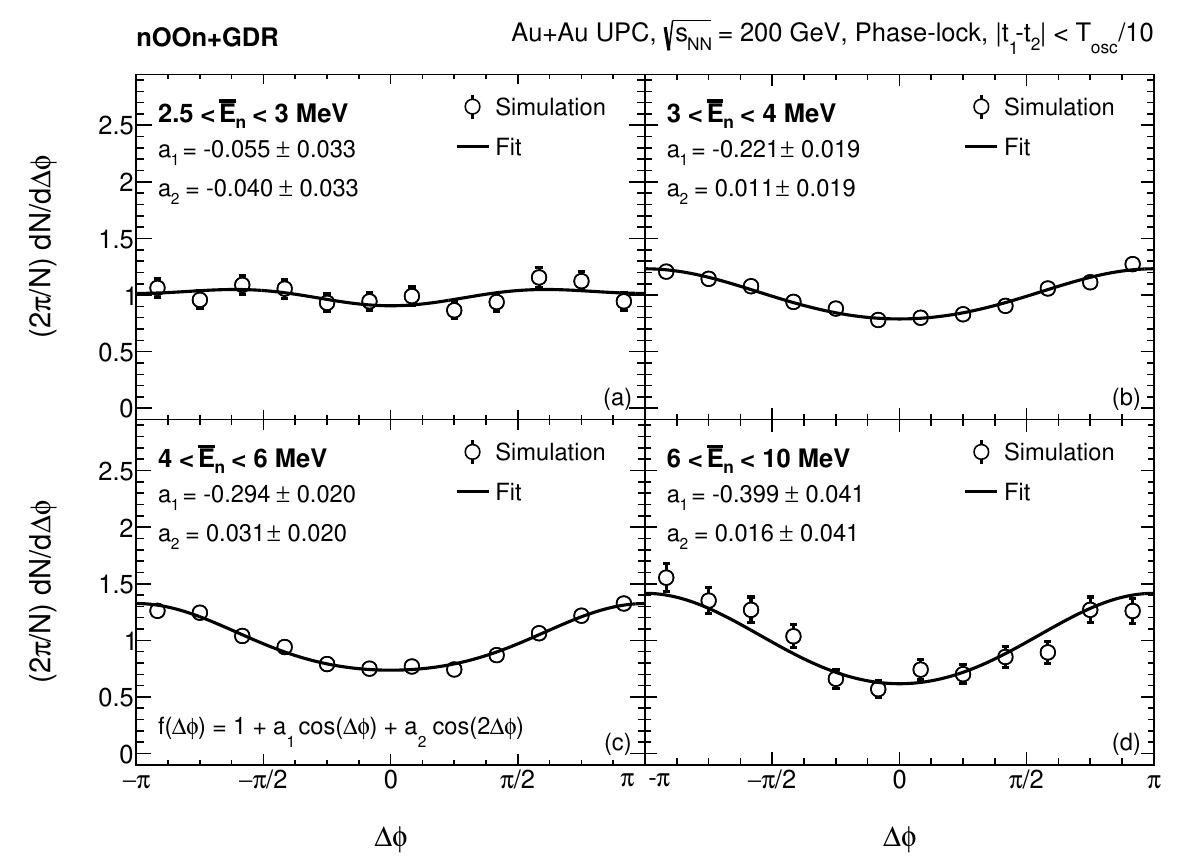}
\caption{Two-neutron relative-azimuth distributions, $dN/d\Delta\phi$, for the time-selected correlated-phase benchmark, containing multiple neutron-multiplicity classes, in Au+Au UPCs at $\sqrt{s_{NN}}=200$ GeV for four intervals of pair-mean neutron kinetic energy $\overline{E}_n=(E_{n,1}+E_{n,2})/2$ in the respective parent-nucleus rest frames, with $|t_1-t_2|<T_{\rm osc}/10$: (a) $2.5<E_n<3$ MeV, (b) $3<E_n<4$ MeV, (c) $4<E_n<6$ MeV, and (d) $6<E_n<10$ MeV. The simulated distributions are fitted with the Fourier expansion $f(\Delta\phi)=1+a_1\cos\Delta\phi+a_2\cos2\Delta\phi$. The extracted values of $a_1$ are summarized as a function of neutron kinetic energy in Fig.~\ref{fig4:a1_en}.
}
  \label{fig3:delphi_en}
\end{figure}

The energy dependence of the phase-lock signal is presented in Fig.~\ref{fig3:delphi_en}. The two-neutron relative-azimuth distributions are shown for four neutron kinetic-energy intervals: (a) $2.5<E_n<3$ MeV, (b) $3<E_n<4$ MeV, (c) $4<E_n<6$ MeV, and (d) $6<E_n<10$ MeV. A clear monotonic enhancement of the dipolar correlation is observed as the neutron energy increases. Near threshold ($2.5<E_n<3$ MeV), the phase-lock signal is strongly suppressed ($a_1=-0.055\pm0.033$), whereas it reaches $a_1=-0.399\pm0.040$ in the highest-energy interval ($6<E_n<10$ MeV).

\begin{figure}[!t]
  \centering
  \includegraphics[width=0.55\linewidth]{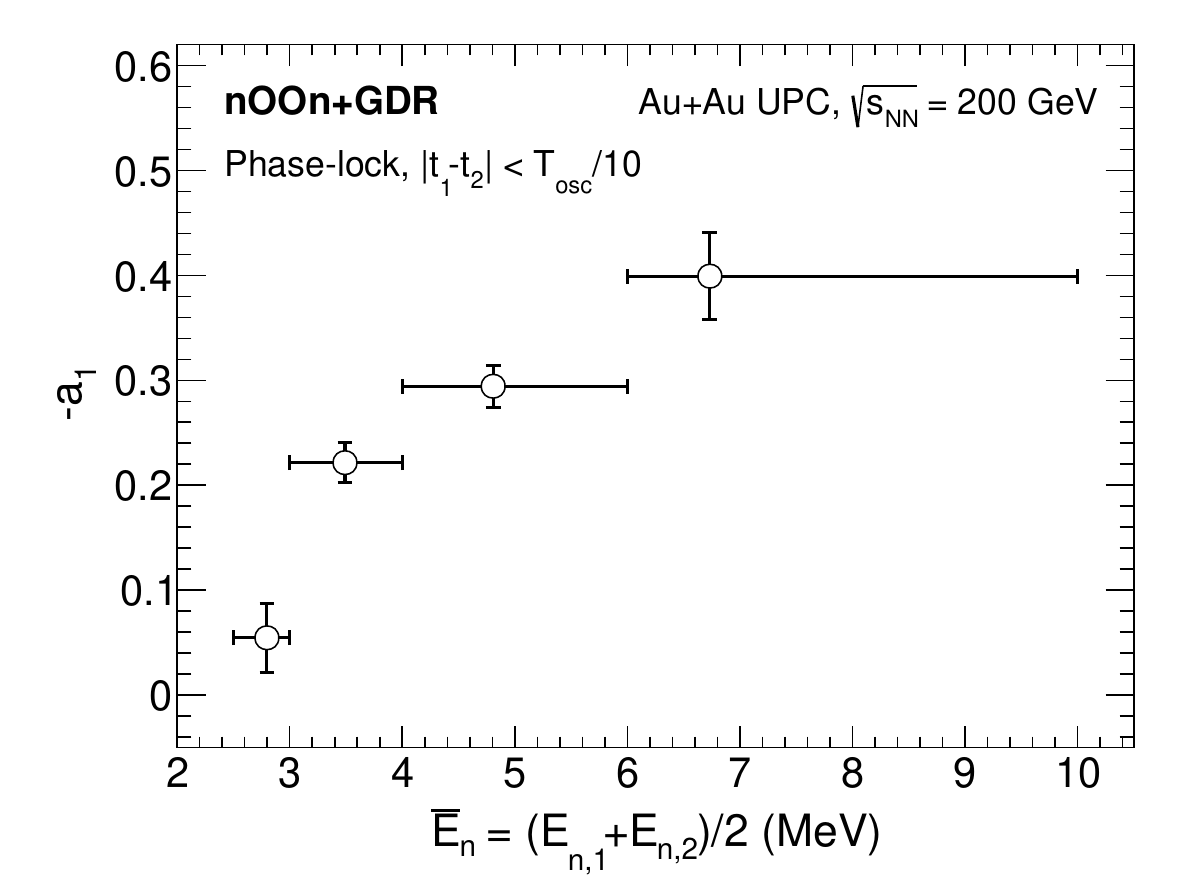}
\caption{Magnitude of the dipolar Fourier coefficient, $-a_1$, versus pair-mean neutron kinetic energy, $\overline{E}_n$, in the respective parent-nucleus rest frames, for the sample selected by $|t_1-t_2|<T_{\rm osc}/10$. The data points are obtained from the Fourier fits to the relative-azimuth distributions shown in Fig.~\ref{fig3:delphi_en}. 
}
  \label{fig4:a1_en}
\end{figure}

Figure~\ref{fig4:a1_en} summarizes this trend by plotting the extracted dipolar coefficient as a function of neutron kinetic energy. The magnitude of $-a_1$ increases from approximately 0.05 to 0.40 in the highest interval, indicating that the phase-lock signal becomes progressively stronger for higher-energy neutrons. This energy dependence admits a natural physical interpretation based on polarized $(\gamma,n)$ measurements in the GDR region. Horikawa \textit{et al.}~\cite{Horikawa2014} observed that higher-energy neutrons exhibit a stronger azimuthal anisotropy than lower-energy neutrons, attributing this behavior to the increasing contribution of prompt direct neutron emission, whereas lower-energy neutrons originate predominantly from statistical cascade and evaporative processes that dilute the polarization asymmetry. In our simulation, neutron emission times are sampled independently of neutron kinetic energy. The energy dependence of the selected first harmonic instead follows the prescribed energy-dependent neutron angular-amplitude model. Photonuclear data motivate a possible relationship between neutron energy and the prompt component, but that relationship is not dynamically implemented here; experimental enrichment of small relative emission-time differences remains to be established.

The same physical picture naturally extends to higher neutron multiplicities ($XnXn$ channels). These events preferentially sample higher photon energies and multi-photon excitations, thereby enhancing the prompt direct GDR component that carries both the polarization-induced quadrupole and the phase-lock dipole. At the same time, the higher excitation energy also increases the contribution from late-time evaporative neutron emission~\cite{Miyamoto:2018PFR,Takeshita:2026xvn}, which is nearly isotropic and therefore dilutes, but does not eliminate, the anisotropic signal. The observed azimuthal correlation is thus a weighted combination of an early, phase-sensitive direct component and a later statistical evaporation background.

Taken together, Figs.~\ref{fig2:delphi}--\ref{fig4:a1_en} demonstrate that the neutron relative-azimuth distribution provides a sensitive discriminator among the three mechanisms discussed in Sec.~\ref{theory}. Electromagnetic recoil generates only a weak dipolar modulation, independent polarized GDR decays produce a quadrupolar baseline, whereas the phase-lock scenario gives rise to a pronounced first-harmonic modulation. In these illustrative configurations, the selected phase-lock sample exhibits a larger dipole than the reference samples. Its experimental distinguishability requires time-inclusive, matched-baseline predictions and detector-response studies. Experimentally, the neutron azimuth is reconstructed from the transverse shower position in the ZDC shower-maximum detectors~\cite{crawford6star,STAR:2005btp}; their finite position resolution reduces each measured coefficient $a_n$ by the product of the azimuthal resolution factors $\langle\cos n(\phi^{\rm meas}-\phi)\rangle$ of the two neutrons. 

\section{Conclusion}

We have investigated the azimuthal correlations of forward neutrons produced in mutual Giant Dipole Resonance (GDR) excitation in ultra-peripheral heavy-ion collisions. Building upon the conventional picture of statistically independent GDR decays, we developed a phase-sensitive GDR decay model and implemented it in a modified version of the $n_{0}^{0}n$ Monte Carlo generator. This framework allows the relative contributions from three distinct mechanisms to be quantified: the weak dipolar modulation arising from electromagnetic recoil, the quadrupolar correlation generated by independent polarized GDR decays, and the phase-sensitive dipolar modulation associated with the phase-lock scenario.

The simulations demonstrate that the three mechanisms lead to qualitatively different signatures in the two-neutron relative-azimuth distribution. Electromagnetic recoil produces only a weak first-harmonic modulation, while independent polarized GDR decays generate a robust quadrupolar baseline with essentially no dipolar component. In contrast, the time-selected phase-lock benchmark gives a pronounced first harmonic. Its magnitude is substantially reduced after averaging over all generated emission-time differences; a quantitative experimental comparison requires matched event selections for the reference mechanisms.

For the time-selected benchmark, the dipolar correlation increases in magnitude with pair-mean neutron kinetic energy through the prescribed angular-anisotropy model. The present simulation does not couple neutron energy to emission time; assessing energy-based experimental selections will require additional modeling.

Although the correlated phase is introduced phenomenologically in the present work, the predicted first-harmonic modulation constitutes a distinctive observable that can be tested experimentally with position-sensitive ZDCs at RHIC and the LHC. Future measurements of neutron relative-azimuth correlations may constrain such phase-sensitive scenarios, provided that accessible selections, conventional contributions, and detector effects are quantitatively accounted for. While observation of such phase-sensitive correlations would not by itself establish entanglement, it could provide a basis for future tests of quantum entanglement between spatially separated GDR excitations.

\section*{ACKNOWLEDGEMENT}
This work is supported in part by the U.S. DOE Office of Science under contracts DE-FG02-89ER40531, DE-SC0012704, and the National Natural Science Foundation of China under Grants No. 12275053. A.A. acknowledges financial support from the Deanship of Scientific Research at Northern Border University through Project No. NBU-SAFIR-2026.

\bibliographystyle{unsrt}
\bibliography{references}
\end{document}